\documentclass[prb,aps,twocolumn,nobibnotes,nofootinbib]{revtex4}
\usepackage{amsmath}
\usepackage{amssymb}
\usepackage{bm}
\usepackage{epsfig}
\usepackage{graphicx}

\newcommand{\divv}{\mathop{\rm div}\nolimits}
\newcommand{\grad}{\mathop{\rm grad}\nolimits}

\newcommand{\res}{\mathop{\rm res}}
\begin{document}

\title{Light-matter interaction in doped microcavities}

\author{N.S.~Averkiev and M.M.~Glazov}
\affiliation{A.F.Ioffe Physico-Technical Institute, Russian
Academy of Sciences, 194021 St.-Petersburg, Russia.}

\begin{abstract}
We discuss theoretically the light-matter coupling in a
microcavity containing a quantum well with a two-dimensional
electron gas. The high density limit where the bound exciton
states are absent is considered. The matrix element of interband
optical absorbtion demonstrates the Mahan singularity due to
strong Coulomb effect between the electrons and a photocreated
hole. We extend the non-local dielectric response theory to
calculate the quantum well reflection and transmission
coefficients, as well as the microcavity transmission spectra. The
new eigenmodes of the system are discussed. Their implications for
the steady state and time resolved spectroscopy experiments are
analyzed.
\end{abstract}

\date{\today} \maketitle

\section{Introduction}

The phenomena of coherent energy transfer between a microcavity
photon and a quantum well exciton have been demonstrated
experimentally for the first time in Ref.~\onlinecite{weisbuch}.
Since when the linear and non-linear effects in quantum
microcavities are extensively studied both experimentally and
theoretically.~\cite{deveaud_book,kavokin03b}

The strong light-matter interaction manifests itself as the
appearance of new eigen-modes, \emph{exciton-polaritons}, being
composite half-light -- half-matter particles. The formation of
the exciton-polaritons can be understood in the terms of two
coupled oscillators describing the confined photon and exciton
respectively. If the coupling constant $V_R$ exceeds the total
damping rate (caused by the photon leakage through the mirrors and
the exciton inhomogeneous broadening) two new eignemodes splitted
by $2V_R$ appear. The microcavity transmission and reflection
coefficients as the functions of the excitation energy and the
detuning between exciton and photon energies show an anticrossing
behavior familiar in the two-level interacting systems, which is
the signature of the \emph{strong-coupling regime}. In
time-resolved experiments the coherent beats of emission are
observed demonstrating the energy transfer between an exciton and
a photon.~\cite{rabi}

The weak doping of the microcavity with electrons is shown to
increase the scattering rates of the
exciton-polaritons.~\cite{kavokin03b,bloch} In the moderate doping
conditions where the quantum well absorbtion edge is determined by
charged excitons ($X^+$ or $X^-$ trions), the strong coupling
between photon and trion was demonstrated.~\cite{Qarry03} The
situation is expected to be drastically different in the highly
doped systems. In this case bound exciton states are known to
vanish due to both the screening of the Coulomb potential and the
state-filling effects, an interband absorbtion is governed by the
Mahan singularity.~\cite{mahan67,Nozieres69a,Hawrylak91} If such a
quantum well is embedded into the microcavity, the confined photon
interacts with a continuum of states which should lead to the
strong differences with the two-oscillator model.

In the present paper we consider the latter regime of light-matter
coupling. We calculate the optical susceptibility of a
two-dimensional electron gas confined in a quantum well with
allowance for the interaction between Fermi-sea of electrons and a
photocreated hole. Further, we use the non-local dielectric
response theory in order to find the quantum well reflection and
transmission coefficients. Then the transfer matrix method is
applied to study the steady-state and time-domain responses of a
microcavity with a doped quantum well. The results are interpreted
in the framework of the simple quantum-mechanical model which
describes the coupling between a discrete state (photon) and a
continuum of electron-hole excitations.

The paper is organized as follows: an expression for the optical
susceptibility of a two-dimensional electron gas with allowance
for the Fermi-edge singularity is derived in Sec. II. Section III
is devoted to the calculation of a quantum microcavity
transmission coefficient. The implications of the Fermi-edge
singularity on time-resolved response of the microcavity is
discussed in Sec. IV.

\section{Susceptibility of a doped quantum well}

Let us consider a doped quantum well made of a direct band
semiconductor with the effective band gap (including quantum well
size-quantization energy) $E_g$. We assume that the quantum well
contains a two-dimensional electron gas with the concentration
$N=k_F^2/2\pi$, where $k_F$ is the Fermi wave vector. The gas
parameter $r_s = \sqrt{2}me^2/(\kappa_0 \hbar^2 k_F)$, where $m$
is electron effective mass, $e$ is the elementary charge and
$\kappa_0$ is the static dielectric constant, characterizes the
ratio between the Coulomb interaction energy of electrons and
their Fermi energy $E_F = \hbar^2 k_F^2/2m$. Gas parameter is
assumed to be the small parameter of our theory, $r_s\ll 1$, i.e.
the Coulomb interaction of the Fermi-level electrons is
negligible.

The absorbtion of a photon with an energy $\hbar \omega$ in the
vicinity of $E_g + E_F$ creates an extra electron in the
conduction band and a hole in the valence band. For simplicity we
assume that the hole is infinitely heavy and interacts with
electrons by a short range attractive potential $V(\bm r) = V_0
\delta(\bm r)$ ($V_0<0$). Latter assumption is violated for the
electrons with the wavevectors $k \ll k_F$ which form bound states
with the hole and screen the long-range Coulomb potential. These
bound states, however, have an energy close to $E_g$ and do not
determine the optical properties of the quantum well for $\hbar
\omega \approx E_g + E_F$. In this spectral range the interband
matrix element of the optical absorbtion is strongly modified due
to the combined effect of the electron-hole interaction and the
step-like change of the density of available states
as~\cite{mahan67,Nozieres69a,Hawrylak91,pardee73,penn81}
\begin{equation}\label{mahan}
M_k = M^0_k \left(\frac{E_F}{\hbar \omega - E_g -
E_F}\right)^{\frac{\delta}{\pi}}.
\end{equation}
Here $k=\sqrt{2m(\hbar \omega - E_g)/\hbar^2}$ is the photoexcited
electron wavevector, $M^0_k$ is the bare matrix element
(calculated without electron-hole interaction) and the power of
singularity $\delta$ is given for the short-range potential by
\begin{equation}\label{delta}
\delta = \delta_0 - \frac{ \delta_0^2}{\pi},
\end{equation}
where $\delta_0=-mV_0/2\hbar^2$ is the scattering phase-shift
introduced by a hole in $s$-channel.~\cite{pardee73,anderson67}
The first term in Eq. \eqref{delta} describes an excitonic
enhancement of the absorbtion and the second term takes into
account the `orthogonality catastrophe': due to the interaction
with hole the electron wavefunctions rearrange and become nearly
orthogonal to the initial wavefunctions. This expression
\eqref{mahan} is valid for $|\hbar \omega - E_g - E_F| \lesssim
E_F$, otherwise $M_k \approx M^0_k \approx const$. An extra factor
of the order of unity  as well as the shift of the absorbtion edge
may appear in Eq. \eqref{mahan} in a more elaborate
approach.~\cite{penn81} We disregard it for the purposes of the
present paper.

The light matter interaction in the planar systems can be most
conveniently described by the non-local dielectric response theory
where the quantum well polarization $\bm P$, caused by the
electromagnetic field $\bm E$ incident along the quantum well
growth axis $z$, can be represented in the integral
form~\cite{andreani91,kavokin03b,ivchenko05a}
\begin{equation}\label{polarization}
\bm P(z) = \int \pi(\omega, z, z') \bm E(z') \mathrm d z',
\end{equation}
where $\pi(\omega, z, z')$ is so-called non-local susceptibility.
It can be written as a sum over all allowed transitions, i.e.
\begin{equation}\label{pi}
\pi(\omega, z, z') = \frac{1}{4}\hbar \kappa_b \omega_{LT}a_B^3
\Phi(z) \Phi(z') G(\omega),
\end{equation}
\[
G(\omega) = \frac{1}{\mathcal S}\sum_{\bm k}
\left|\frac{M_k}{M_k^0}\right|^2 \frac{1- n_F(k)}{E_g +
\frac{\hbar k^2}{2m} - \hbar \omega -\mathrm i 0} ,
\]
where $\mathcal S$ is the normalization area, $\kappa_b$ is the
background high-frequency dielectric constant, $a_B$ and
$\omega_{LT}=4\hbar e^2p_{cv}^2/(E_g^2 m_0^2 \kappa_b a_B^3)$ are
the Bohr radius and longitudinal-transverse splitting of the bulk
exciton, $p_{cv}$ is the interband dipole matrix element,
$\Phi(z)=\varphi_e(z)\varphi_h(z)$ is the electron-hole pair
envelope along the growth axis, and $n_F(k)=1$ for $k\leqslant
k_F$ and $0$ otherwise is the Fermi distribution function. In Eq.
\eqref{pi} we assumed that the quantum well is thin enough that
the Coulomb effect on the motion of electron and hole along $z$
axis can be neglected, we disregarded the inhomogeneous broadening
of the electron states, and set temperature to be zero.

Substituting Eq. \eqref{mahan} into Eq. \eqref{pi} one obtains a
closed form expression for the susceptibility for $|\hbar \omega -
E_g - E_F| \lesssim E_F$ and $\delta>0$:
\begin{equation}\label{pi_mah}
G(\omega) = \frac{m}{2 \hbar^2\sin{2\delta}} \left\{
\begin{array}{cc}
\left(-\mathcal E\right)^{-\frac{2\delta}{\pi}}, &
\mathcal E<0,\\
e^{2\mathrm i \delta} \mathcal E^{-\frac{2\delta}{\pi}}, &
\mathcal E>0.\\
\end{array}\right.
\end{equation}
Here reduced energy $\mathcal E = (\hbar \omega - E_g - E_F)/E_F$.
One can see that below the threshold, $\mathcal E<0$, the
susceptibility is real, while above the threshold, $\mathcal E>0$,
the susceptibility contains both real and imaginary parts. Note
that formally the integral describing the real part of
$\pi(\omega,z,z')$ in \eqref{pi} is logarithmically divergent as
$M_k/ M^0_k\to 1$ for $k\to \infty$. To obtain a finite result
either non-resonant terms in the susceptibility or corrections to
the effective mass method are needed. In the doped system,
however, for $\hbar^2k^2/2m \sim E_F$ the broadening of the single
particle levels due to electron-electron scattering makes large
$k$ contribution negligible and one may use Eq. \eqref{mahan} for
the interband matrix element in a whole relevant spectral range.

\begin{figure}[htb]
\centering
\includegraphics[width=0.75\linewidth]{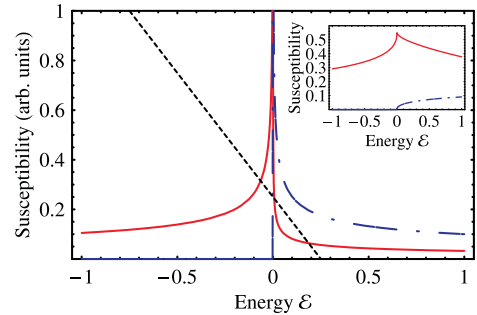}
\caption{(Color online). Susceptibility of the doped quantum well
$G$ [Eq. \eqref{pi_mah}], plotted as a function of $\mathcal
E=(\hbar \omega - E_g - E_F)/E_F$. Real part is shown by red solid
line, blue dash-dot line represents the imaginary part of $G$.
Black dotted line shows $\hbar\omega_c/E_F - \mathcal E$ ($\hbar
\omega_c/E_F = 0.25$) and the intersection points are the
solutions of Eq. \eqref{eigenmodes2}. The parameters used are:
$\delta = 0.63$, dimensionless light-matter coupling constant
$\mathcal C = 0.1$. An inset shows susceptibility calculated for
negative value of $\delta=-0.63$.} \label{fig:suscept}
\end{figure}

If the enhancement power $\delta=0$ then the power-law singularity
in susceptibility is replaced by the logarithmic one, $G(\omega)
\propto \ln{(-\tilde{\mathcal E} /\mathcal E)}$, where
$\tilde{\mathcal E}$ is the cut-off energy.~\cite{ivchenkopikus}
In the case of negative power $\delta<0$ in Eq. \eqref{mahan} the
singularity is absent
\begin{equation}\label{pi_mah_no_sing}
G(\omega) = \frac{m}{2 \hbar^2} \left\{
\begin{array}{cc}
\Gamma{\left(1-\frac{2\delta}{\pi}\right)}
\Gamma{\left(\frac{2\delta}{\pi}, -\frac{\mathcal
E}{\tilde{\mathcal E}}\right)} \left(-\mathcal
E\right)^{-\frac{2\delta}{\pi}}, &
\mathcal E<0,\\
\Re{\left\{\Gamma{\left(1-\frac{2\delta}{\pi}\right)}
\Gamma{\left(\frac{2\delta}{\pi}, -\frac{\mathcal
E}{\tilde{\mathcal E}}\right)} \left(-\mathcal
E\right)^{-\frac{2\delta}{\pi}} \right\} } + \mathrm i \mathcal
E^{-\frac{2\delta}{\pi}}, &
\mathcal E>0,\\
\end{array}\right.
\end{equation}
where $\Gamma(a,b)$ is incomplete gamma-function, $\tilde{\mathcal
E}$ is the cut-off energy (which we introduce as $M_k \propto
\mathcal E^{-\delta/\pi} e^{-\mathcal E/2\tilde{\mathcal E}}$).
Note that in the vicinity of the threshold $G(\omega)$ remains
finite. From now on we concentrate on the most interesting case of
$\delta>0$ where the singularity is observed; we shortly discuss
the light-matter coupling for negative $\delta$ below.

Figure~\ref{fig:suscept} shows the dimensionless susceptibility
$G$ as a function of $\mathcal E$ in the vicinity of the quantum
well optical absorbtion edge. The parameters used are given in the
caption to the figure. Note, that the value of $\delta\approx
0.63$ is exaggerated in the illustrative purposes, qualitatively
all the results hold for smaller but positive values of $\delta$
as well. One can see that the imaginary part (dash-dot line)
appears only for $\mathcal E>0$, i.e. for $\hbar \omega> E_g +
E_F$. The real part (solid) is non-zero for any $\mathcal E$ and
has a sharp asymmetric peak in the vicinity of $\mathcal E=0$. We
note that the real part of $G$ keeps its sign on both sides of the
absorbtion edge being in a constrast with the case of a single
exciton resonance where $G(\omega) \propto (\omega_{exc} -
\omega)^{-1}$ and changes its sign at exciton resonance frequency
$\omega_{exc}$. For the sake of comparison, an inset to
Fig.~\ref{fig:suscept} shows suseptibility calculated in the case
of negative $\delta$: the imaginary part starts from $0$ and real
part takes a finite value at the threshold.

\section{Light reflection and transmission}

In order to calculate the optical transmission of a microcavity
containing a doped quantum well we first have to find the quantum
well reflection and transmission coefficients, then to apply the
transfer matrix method to analyze the microcavity transmission and
new eigenmodes which arise due to the light-matter interaction.

We write Maxwell equation for the electric field vector $\bm E$
as~\cite{kavokin03b,ivchenko05a}
\begin{equation}\label{maxwell1}
\Delta \bm E + q^2 \bm E = -4\pi \left(\frac{\omega}{c}\right)^2
\left[\bm P + \frac{1}{q^2}\grad\divv \bm P\right],
\end{equation}
where $q$ is the light wavevector $q=\sqrt{\kappa_b}\omega/c$ and
$\bm P$ is the polarization induced by the quantum well, Eq.
\eqref{polarization}. Let us assume that the light is incident
along the normal to the quantum well plane from the negative
direction of $z$-axis. One can show that $\divv \bm E = \divv \bm
P \equiv 0$, thus vector equation \eqref{maxwell1} reduces to the
single equation for the scalar field amplitude $E$ (in-plane
component of $\bm E$). Its general solution corresponding to the
absence of the light incident from $z>0$ can be written
as~\cite{ivchenko05a}
\begin{equation}\label{maxwell:sol}
E(z) = E_0 e^{\mathrm i q z} + 2\pi \mathrm i q^{-1}
\left(\frac{\omega}{c}\right)^2\int \mathrm d z' e^{\mathrm i q
|z-z'|} P(z').
\end{equation}
Here $E_0$ is the incident field amplitude. Eq.
\eqref{maxwell:sol} can be reduced to linear algebraic one by the
multiplication by $\Phi(z)$ and integration over $z$. It allows
immediately to find the quantum well amplitude reflection
coefficient:
\begin{equation}\label{refl_norm}
r = \frac{\mathrm i Q G(\omega)}{1-\mathrm i Q G(\omega)},
\end{equation}
where
\begin{equation}\label{Q}
Q = \frac{q}{2}\hbar \omega_{LT}\pi a_B^3 \left[\int \Phi(z)
e^{\mathrm i q z} \mathrm d z\right]^2.
\end{equation}
Here integration is carried out within the quantum well, i.e. in
the domain where $\Phi(z)$ does not vanish. In Eq.
\eqref{refl_norm} the renormalization of the absorbtion edge
frequency is neglected as it arises to the extent of the small
parameter $q a$, where $a$ is the quantum well width. Within the
same approximation one may omit factor $e^{\mathrm i q z}$ in
Eq.~\eqref{Q}. Amplitude transmission coefficient of the quantum
well reads $t=1+r$. We note that $Q$ is related to the exciton
radiative broadening in the undoped quantum well $\hbar
\Gamma_0$~[Ref. \onlinecite{ivchenko05a}] as $Q  = \hbar
\Gamma_0\phi^{-2}_0$, where $\phi_0$ is the exciton in-plane
relative motion wavefunction taken at coinciding electron and hole
coordinates.

Let us consider a microcavity with a doped quantum well embedded
between two Bragg mirrors. The transfer matrix formalism allows us
to write the microcavity transmission coefficient at normal
incidence in the following compact
form~\cite{kavokin03b,ivchenko05a}
\begin{equation}\label{tqm}
t_{qmc} = \frac{t_m \tilde{t} e^{\mathrm i \psi_b}}{1 - \tilde{r}
r_{m} e^{\mathrm i \psi_b}},
\end{equation}
where $t_m$, $r_m$ are the transmission and reflection
coefficients of Bragg mirrors (we assume that left and right
mirrors are the same), $\psi_b = n_b k_z L_b$ with $L_b$ being the
active layer width, and we have introduced the quantities
\[
\tilde{r} = r + \frac{t^2 r_m e^{\mathrm i \psi_b}}{1 - r r_m
e^{\mathrm i \psi_b}}, \quad \tilde{t} =  \frac{t t_m e^{\mathrm i
\psi_b/2}}{1 - r r_m e^{\mathrm i \psi_b}}.
\]

Below we discuss the steady-state response of the microcavity, the
time-resolved emission will be considered in the next section.

\begin{figure*}[htb]
\centering
\includegraphics[width=0.85\linewidth]{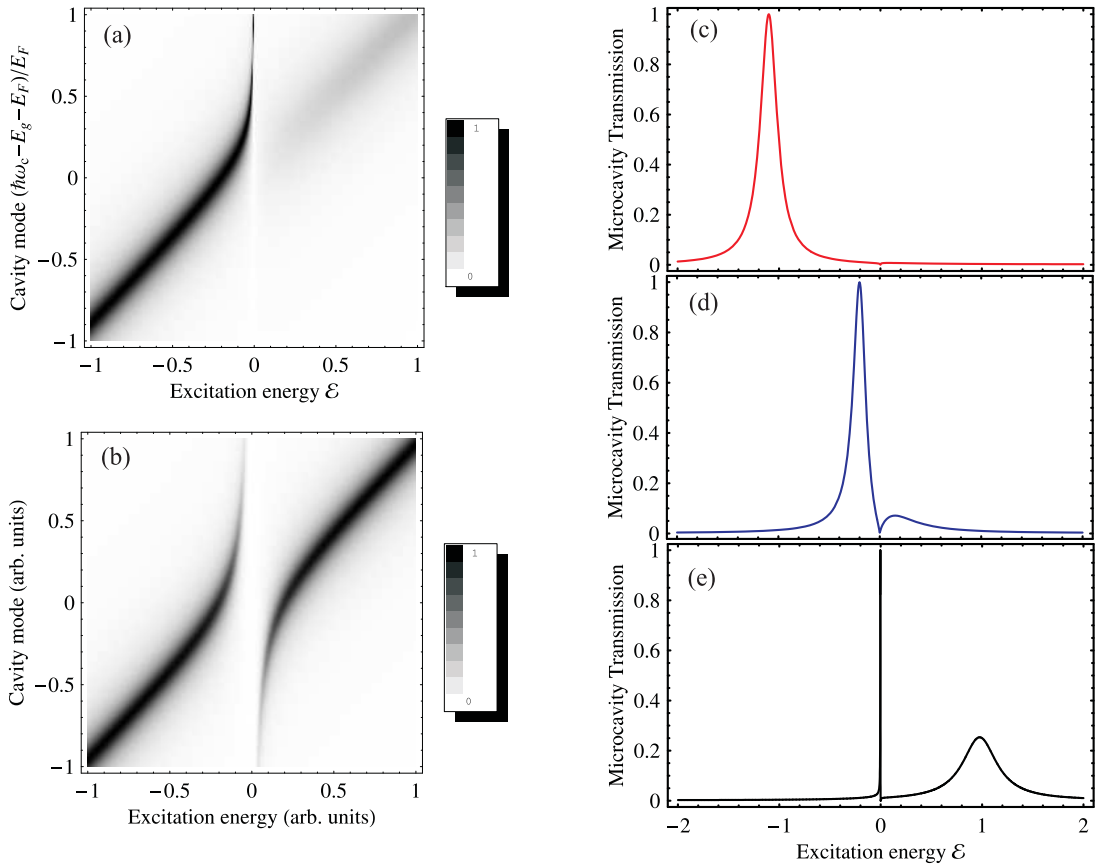}
\caption{(Color online). Quantum microcavity transmission
coefficient $|t_{qmc}|^2$ [Eq. \eqref{tqm}] as a function of the
excitation energy $\mathcal E = (\hbar \omega - E_g - E_F)/E_F$
and the cavity mode energy $(\hbar \omega_c -E_F)/E_F$. (a) and
(b) are the surface plots (the gray-level scale is shown to the
right) corresponding to the Mahan singularity and discrete
excitonic state, respectively. (c), (d) and (e) are the horizontal
cuts of the panel (a) for different cavity mode energies $(\hbar
\omega_c -E_F)/E_F = -1$, $0$ and $1$, from the top to the bottom.
The parameters used are: $\delta = 0.63$, $\mathcal C = 0.1$,
$R=0.98$.} \label{fig:transm}
\end{figure*}

Figure \ref{fig:transm} presents the microcavity transmission
coefficient $|t_{qmc}|^2$ calculated as a function of excitation
energy and cavity mode position. The maximum of the transmission
coefficient [dark area of Fig. \ref{fig:transm}(a)] corresponds to
the eigenmodes of the quantum microcavity. They can be found
analytically as zeros of the denominator in Eq. \eqref{tqm}:
\begin{equation}\label{eigenmodes}
\mathcal D = [r_m(2r+1)e^{\mathrm i \psi_b} - 1](r_m e^{\mathrm i
\psi_b}+1).
\end{equation}
The second factor in Eq. \eqref{eigenmodes} corresponds to the
cavity modes which are not coupled to the two-dimensional electron
gas. The first factor describes new (polaritonic) modes arising
due to the light-matter interaction. Assuming that the relevant
frequencies lie in the vicinity of the stop-band center of Bragg
mirror one may use the following approximate expressions for the
Bragg mirror reflection and transmission
coefficients~\cite{kavokin03b,panzarini99,ivchenko05a}
\begin{equation}\label{rmirrors}
r_m = \sqrt{R} e^{\mathrm i \psi_m(\omega)}, \qquad t_m = \pm
\mathrm i r_m \sqrt{(1-R)/R}.
\end{equation}
Here $R$ is the intensity reflection coefficient and
$\psi_m(\omega)$ is a phase. In the vicinity of the stop-band
center it is a linear function of $\omega$.~\cite{panzarini99} Eq.
\eqref{eigenmodes} clearly shows that the only independent phase
parameter is $\psi = \psi_m+ \psi_b$, which can be recast in form
$\psi = (\omega - \omega_c)/\bar \omega$, where $ \omega_c$ is the
cavity resonance frequency, and $c/\bar\omega = (L_{DBR} + L_b)
n_b$ is the effective cavity length (including the active layer
length, $L_b$, and the mirror penetration length,
$L_{DBR}$)~\cite{kavokin03b}. Substituting  Eq. \eqref{rmirrors}
into the first factor of Eq. \eqref{eigenmodes} under assumption
that $(\omega-\omega_c)/\bar\omega \ll 1$ and $R\to 1$ we arrive
to
\begin{equation}\label{eigenmodes2}
\omega_c - \omega = 2 Q \bar \omega G(\omega).
\end{equation}
It is convenient to introduce the dimensionless coupling constant
$\mathcal C = mQ \bar \omega/(\pi \hbar E_F)$, which can be
related to the coupling constant $V_R$ of the undoped cavity as
$\mathcal C = (k_F a_B V_R)^2/32 E_F^2$.

Equation \eqref{eigenmodes2} has a very transparent physical
meaning: it describes the linear coupling between the microcavity
photon and the continuum of states of all electron-hole pair
exciations. It is equivalent to the eigenenergy equation for the
following Hamiltonian
\begin{equation}\label{hamiltonian}
\mathcal H = \hbar \omega_c c^\dag c + \sum_{\bm k} \left(E_g +
\frac{\hbar^2 k^2}{2m}\right) a_{\bm k}^\dag a_{\bm k} +
\end{equation}
\[
\sqrt{\frac{2 Q\bar\omega \hbar}{\mathcal S}}
 \sum_{\bm k} \left|\frac{M_k}{M_k^0}\right|^2  [1- n_F(k)] (c^\dag
a_{\bm k} + c a^\dag_{\bm k}),
\]
where $c^\dag$, $c$ are the creation (annihilation) operators for
the cavity photon and $a_{\bm k}^\dag$, $a_{\bm k}$ are the
creation (annihilation) operators for electron-hole pair
excitations. Thus, the description of the light matter interaction
in the doped system is reduced to the quantum mechanical problem
of the interaction between the discrete state and a
continuum.~\cite{suris}

The graphical solution of Eq. \eqref{eigenmodes2} is represented
in Fig.~\ref{fig:suscept}. The dotted line shows the left-hand
side of Eq. \eqref{eigenmodes2} and the blue solid line shows the
right hand side. All the states with $\hbar \omega \geq E_g+ E_F$
belong to continuum. Thus, non-trivial solutions of Eq
\eqref{eigenmodes2} can be for $\hbar \omega < E_g + E_F$ only.
One may identify two regimes of light-matter coupling depending on
the number of solutions of Eq. \eqref{eigenmodes2}.

In the first case, the discrete state with an energy $\hbar
\omega_0$ exists for all the values of $\hbar\omega$, i.e.
$G(\omega)$ diverges for $\hbar\omega=E_g + E_F$. This is the case
of the Mahan singularity in the optical absorbtion, Eq.
\eqref{pi_mah}, or of the undoped two-dimensional system. In the
second regime, where $G[(E_g+ E_F)/\hbar]$ is finite, the discrete
state exists only for $\hbar \omega_c<\hbar \omega_m$, where the
limiting frequency $\omega_m=E_g + E_F + 2Q \bar\omega G[(E_g+
E_F)/\hbar]$. This case can be realized if the `optical' density
of states $\propto |M_k|^2$ is zero at the threshold as it takes
place for an undoped crystal in the vicinity of the fundamental
absorbtion edge or for $\delta<0$, corresponding to the dominant
contribution of the orthogonality
catastrophe,~Eq.~\eqref{pi_mah_no_sing}.

In other words, with an increase of the cavity mode frequency
$\omega_c$ the photon state repels from the continuum due to the
light-matter interaction, but depending on the coupling strength
it can either be pushed away from the continuum or can merge a
continuum. It is illustrated in Fig.~\ref{fig:transm}(a): as the
cavity mode position shifts to higher energies, the energy
corresponding to the maximum of transmission increases and
approaches the singularity position $\hbar \omega = E_g+ E_F$. The
strong repulsion of the photonic state from the continuum is seen;
in our case of singular susceptibility ($\delta>0$) the discrete
photonic states survives for any cavity mode position, see panels
(c)--(e). This discrete photonic state peak becomes narrower for
larger $\hbar \omega_c$ [it is most pronounced in
Fig.~\ref{fig:transm}(e)] but the allowance for the inhomogeneous
broadening of the Mahan singularity will smear this peak. Due to
an enhanced absorbtion at $\hbar \omega> E_g +E_F$ the
transmission tends to zero at $\hbar \omega$ approaching to $E_g+
E_F$ from high-energy side. An additional maximum in transmission
appears at $\hbar \omega>E_g+ E_F$ at the energy approximately
equal to bare cavity mode position $\hbar \omega_c$ becoming more
pronounced with an increase of cavity mode energy. It can be
attributed to the resonant state formed by the cavity mode
`inside' the continuum of electron-hole excitations and
corresponds to the third intersection of the solid and dotted
curves in Fig.~\ref{fig:suscept}, see next section.

\begin{figure}[htb]
\centering
\includegraphics[width=0.65\linewidth]{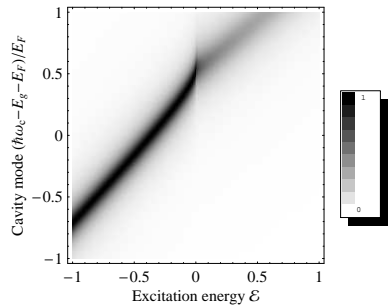}
\caption{Quantum microcavity transmission coefficient
$|t_{qmc}|^2$ [Eq. \eqref{tqm}] as a function of the excitation
energy $\mathcal E = (\hbar \omega - E_g - E_F)/E_F$ and the
cavity mode energy $(\hbar \omega_c -E_F)/E_F$ calculated for the
case of absent singularity. The parameters used are: $\delta =
-0.63$, $\mathcal C = 0.1$, $R=0.98$.} \label{fig:transm:no_sing}
\end{figure}

For the smaller but non-negative values of $\delta$ the calculated
transmission spectra are qualitatively the same. The decrease of
$\delta$ leads to the sharper peak with larger wings of
susceptibility's real part. It results in less pronounced discrete
state and smaller spacing between absorbtion edge and the last
transmission maximum.

The case of the negative $\delta$ is presented in
Fig.~\ref{fig:transm:no_sing}. In this case the discrete `dressed'
photonic state survives only up to $\omega_c = \omega_m$. The
behaviour of the transmission maximum reflects the real part of
susceptibility and a peak is seen in the vicinity of the
absorbtion edge. The absorbtion above the edge is smaller as
compared to the case of $\delta>0$ thus resonant states inside the
continuum are more pronounced.

This situation depicted in Figs.~\ref{fig:transm}(a),
\ref{fig:transm:no_sing} is qualitatively different from that
observed in undoped cavities tuned to the vicinity of exciton
resonance, see Fig.~\ref{fig:transm}(b). In this case cavity
photon and exciton can be described as two oscillators. Their
coupling leads to the repulsion and anticrossing of the levels, in
the strong coupling regime two discrete states (polaritons)
corresponding to the maxima of the transmission are present.

\section{Time-resolved emission}

The light-matter interaction manifests itself in the time resolved
experiments as well. Consider an experiment where the bare photon
mode is excited by a short pulse and the emission intensity is
analyzed. In the case of the undoped quantum well and a
microcavity tuned to the exciton resonance this intensity
demonstrates the quantum beats, corresponding to the energy
transfer between the photon and exciton states.~\cite{rabi} In the
case of doped system the photon interacts with a continuum of
states and the time-domain response is drastically different.

\begin{figure}[htb]
\centering
\includegraphics[width=0.75\linewidth]{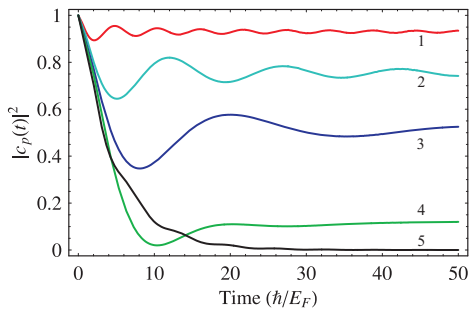}
\caption{(Color online). The photon intensity $|c_{p}(t)|^2$ as a
function of time calculated according to Eq. \eqref{result}.
Different curves correspond to different cavity mode energies
$(\hbar \omega_c -E_F)/E_F = -1$, $-0.25$, $0$, $0.5$ and $1$
(curves 1--5, respectively). The parameters used are: $\delta =
0.63$, $\mathcal C = 0.1$.} \label{fig:cphot}
\end{figure}

In order to find the time-resolved emission from the microcavity
containing a doped quantum well we solve Schroedinger equation
with Hamiltoninan~\eqref{hamiltonian} and represent the
wavefunction of the system as
\begin{equation}\label{psi}
\Psi(t)  = \left[c_p(t) c^\dag + \sum_{\bm k} c_{\bm k} (t)
a^\dag_{\bm k}\right] |0\rangle.
\end{equation}
Here $|0\rangle$ is the `vacuum state', $c_p(t)$ and $c_{\bm
k}(t)$ are time-dependent coefficients of the photon and
continuums states. Substituting Eq. \eqref{psi} into the temporal
Schroedinger equation
\[ \mathrm i \hbar \frac{\partial
\Psi(t)}{\partial t} = \mathcal H \Psi(t),
\]
one arrives to the system of linear coupled equations for the
functions $c_p(t)$, $c_{\bm k}(t)$. It can be solved by Laplace
transformation.

We consider the most interesting case of the initial excitation of
the bare photon mode $[c_p(0)=1]$ and analyze the time dependence
$c_p(t)$ which corresponds to the photon fraction of the
polaritonic mode (i.e. quantum microcavity emission amplitude):
\begin{equation}\label{bromwich}
c_p(t) =  \int_{\gamma-\mathrm i \infty}^{\gamma+\mathrm i \infty}
\frac{e^{st}}{ s +\mathrm i \omega_c - 2 \mathrm i Q \bar \omega
G(\mathrm i\hbar s)} \frac{\mathrm d \hbar s}{2\pi\mathrm i},
\end{equation}
where $\gamma$ is sufficiently large real number. We note that
$\mathrm i 0$ term in the denominator of $G(\omega)$ should be
omitted in Eq. \eqref{bromwich}.

This integral can be calculated by the standard
procedure.~\cite{kyrola} Note that the subintegral expression has
following singularities. First, there may be a pole for the
imaginary $s = -\mathrm i \omega_0$, there $\omega=\omega_0$
satisfies the bound-state equation \eqref{eigenmodes2} which
corresponds to the discrete state. Second, there is a branch-cut
along the positive part of the imaginary axis $E_g + E_F \leq
-\mathrm i s$ which corresponds to the contribution of the
continuum. As a result, integral \eqref{bromwich} is recast in a
form
\begin{equation}\label{result}
c_p(t) =  \res_{\omega = \omega_0}{\frac{ e^{-\mathrm i \omega_0
t}}{\omega - \omega_c + 2 Q \bar \omega G(\omega)}} +
\end{equation}
\[
\int_{E_g +E_F}^{\infty} \frac{e^{-\mathrm i \tilde st}
\alpha^2(\hbar \tilde s) \mathrm d \hbar \tilde s}{\left[\hbar
\tilde s - \hbar \omega_c + vp\int_{E_g +E_F}^{\infty}
\frac{\alpha^2(E)}{E-\hbar \tilde s} \mathrm d E\right]^2 +
\pi^2\alpha^4(\tilde s)},
\]
where
\[
\alpha(E) =  \frac{m Q\bar\omega}{\pi\hbar^2}
\left|\frac{M_k}{M_k^0}\right|^2, \quad
k=\sqrt{\frac{2m(E-E_g)}{\hbar^2}}.
\]

This result can be interpreted in the terms of eigenmodes of the
system. The first term in Eq. \eqref{result} corresponds to the
discrete state (i.e. ``dressed photon'') contribution, while the
second term describes the effects of the continuum. Initially
excited bare photonic mode is divided among new eigenmodes:
discrete state (if it exists) and a continuum. The part
corresponding to the discrete state survives, while that of
continuum is spread over all possible energies. The main
contribution to the second term of Eq. \eqref{result} comes from
the points $\tilde s_i$ where the subintegral expression has a
sharp maximum (i.e. to the intersection points for $\mathcal E>0$
in Fig.~\ref{fig:suscept}). In a crude approximation each of such
contributions reads
\[
\exp{[-\mathrm i \tilde s_i t - \alpha^2(\hbar \tilde
s_i)\hbar^{-1} t]},
\]
thus it can be considered as a quasi-bound state (or resonance).
The microcavity emission $\propto |c_p(t)|^2$ will demonstrate the
decaying quantum beats between the discrete state (dressed photon)
and the continuum (resonances),
\begin{equation}\label{beats_damp}
|c_{p}(t)|^2 \propto \sum_i \cos{[(\omega_0 - \tilde s_i)t +
\varphi_i]}\exp{[-\alpha^2(\hbar \tilde s_i)\hbar^{-1} t]},
\end{equation}
where $\varphi_i$ are the initial phases of the beats.

Fig. \ref{fig:cphot} shows the results of numerical integration in
Eq. \eqref{result}. The decaying beats are observed. The beats
frequency is a non-monotonous function of detuning: it decreases
as $\hbar \omega_c$ approaches $E_g+E_F$ from below (as the main
contribution to beats are given by the first resonance near the
Mahan singularity, see first intersection point in
Fig.~\ref{fig:suscept}), reaches a minimum at $\hbar \omega_c
\approx E_g+E_F$ and then increases, as the energy separation
between the discrete state (with energy close to $E_g+E_F$) and
the second resonance increases. The damping of the beats results
from the non-commensurability of the continuum frequencies.

In this treatment we have disregarded the photon leakage through
the Bragg mirrors and inhomogeneous broadening of the Fermi-edge
singularity. Both of these effects give rise to an additional
damping of the beats as well as to the overall decay of
$c_{p}(t)$.

The time-domain response of the microcavity in the case where the
singularity is absent, $\delta<0$ in Eq. \eqref{mahan} is
qualitatively the same. In the case of $\omega_c < \omega_m$ the
real discrete photonic state exists and the emission intensity is
similar to that presented in Fig.~\ref{fig:cphot}. For $\omega_c >
\omega_m$ only contribution of continuum remains and $c_p(t)$
decays to zero even without photon leakage through the mirrors and
inhomogeneous broadening.

\section{Conclusions}

To conclude, we have theoretically studied the light-matter
coupling in the microcavities containing doped quantum wells. In
such structures the cavity photon interacts with the continuum of
the electron-hole pair excitations. The new eigenmodes of the
system are the discrete state (corresponding to the `dressed'
cavity photon) and the continuum of electron-hole pairs modified
by the interaction with the cavity photon. We have analyzed the
effects of the `optical' density of states on the bound state
formation. It was shown that the step-like or singular behavior of
the absorbtion coefficient at the threshold implies the bound
state presence for any detuning between the cavity mode and the
threshold energy. If the optical density of states is non-singular
at the threshold, the bound photonic state survives only up to
some limiting detuning. We have investigated the reflection and
transmission spectra of the doped microcavities and shown that
they are qualitatively different from those obtained in the
undoped case where discrete excitonic state is coupled to the
discrete photonic state. The time domain response of the doped
microcavity is shown to demonstrate the damped beats revealing the
energy transfer between the discrete state and the continuum and
the energy spread in the continuum.

\begin{acknowledgments}
  We acknowledge the financial support
  by RFBR, Programs of RAS and ``Dynasty'' foundation---ICFPM.
\end{acknowledgments}


\begin{thebibliography}{11}
\expandafter\ifx\csname
natexlab\endcsname\relax\def\natexlab#1{#1}\fi
\expandafter\ifx\csname bibnamefont\endcsname\relax
  \def\bibnamefont#1{#1}\fi
\expandafter\ifx\csname bibfnamefont\endcsname\relax
  \def\bibfnamefont#1{#1}\fi
\expandafter\ifx\csname citenamefont\endcsname\relax
  \def\citenamefont#1{#1}\fi
\expandafter\ifx\csname url\endcsname\relax
  \def\url#1{\texttt{#1}}\fi
\expandafter\ifx\csname
urlprefix\endcsname\relax\def\urlprefix{URL }\fi
\providecommand{\bibinfo}[2]{#2}
\providecommand{\eprint}[2][]{\url{#2}}

\bibitem{weisbuch} C. Weisbuch, M. Nishioka, A. Ishikawa, and Y.
Arakawa, Phys. Rev. Lett. {\bf 69}, 3314 (1992).

\bibitem[{\citenamefont{Kavokin and Malpuech}(2003)}]{kavokin03b}
\bibinfo{author}{\bibfnamefont{A.}~\bibnamefont{Kavokin}} \bibnamefont{and}
  \bibinfo{author}{\bibfnamefont{G.}~\bibnamefont{Malpuech}},
  \emph{\bibinfo{title}{Cavity Polaritons}}, vol.~\bibinfo{volume}{32} of
  \emph{\bibinfo{series}{Thin Films and Nanostructures}}
  (\bibinfo{publisher}{Elsevier}, \bibinfo{year}{2003}).

\bibitem{deveaud_book} Deveaud, Benoit (Ed.) \textit{The Physics of Semiconductor Microcavities
From Fundamentals to Nanoscale Devices}, Wiley-VCH (2006).



\bibitem{rabi} S. Jiang, S. Machida, Y. Takiguchi, and Y.
Yamamoto, Appl. Phys. Lett. {\bf 73}, 3031 (1998).

\bibitem{bloch} D. Bajoni, M. Perrin, P. Senellart, A. Lema$\hat{\mbox{i}}$tre,
 B. Sermage, and J. Bloch, Phys. Rev. B {\bf 73}, 205344 (2006).

\bibitem[{\citenamefont{Qarry et~al.}(2003)\citenamefont{Qarry, Rapaport,
  Ramon, Cohen, Ron, and Pfeiffer}}]{Qarry03}
\bibinfo{author}{\bibfnamefont{A.}~\bibnamefont{Qarry}},
  \bibinfo{author}{\bibfnamefont{R.}~\bibnamefont{Rapaport}},
  \bibinfo{author}{\bibfnamefont{G.}~\bibnamefont{Ramon}},
  \bibinfo{author}{\bibfnamefont{E.}~\bibnamefont{Cohen}},
  \bibinfo{author}{\bibfnamefont{A.}~\bibnamefont{Ron}}, \bibnamefont{and}
  \bibinfo{author}{\bibfnamefont{L.~N.} \bibnamefont{Pfeiffer}},
  \bibinfo{journal}{Semicond. Sci. and Tech.}
  \textbf{\bibinfo{volume}{18}}, \bibinfo{pages}{S331} (\bibinfo{year}{2003}).

\bibitem[{\citenamefont{Mahan}(1967{\natexlab{a}})}]{mahan67}
\bibinfo{author}{\bibfnamefont{G.~D.} \bibnamefont{Mahan}},
  \bibinfo{journal}{Phys. Rev.} \textbf{\bibinfo{volume}{153}},
  \bibinfo{pages}{882} (\bibinfo{year}{1967}{\natexlab{a}});
\bibinfo{author}{\bibfnamefont{G.~D.} \bibnamefont{Mahan}},
  \bibinfo{journal}{Phys. Rev.} \textbf{\bibinfo{volume}{163}},
  \bibinfo{pages}{612} (\bibinfo{year}{1967}{\natexlab{b}}).

\bibitem[{\citenamefont{ROULET et~al.}(1969)\citenamefont{ROULET, GAVORET, and
  NOZI\`ERES}}]{Nozieres69a}
\bibinfo{author}{\bibfnamefont{B.}~\bibnamefont{Roulet}},
  \bibinfo{author}{\bibfnamefont{J.}~\bibnamefont{Gavoret}}, \bibnamefont{and}
  \bibinfo{author}{\bibfnamefont{P.}~\bibnamefont{Nozi\`eres}},
  \bibinfo{journal}{Phys. Rev.} \textbf{\bibinfo{volume}{178}},
  \bibinfo{pages}{1072} (\bibinfo{year}{1969}).

\bibitem[{\citenamefont{Hawrylak}(1991)}]{Hawrylak91}
\bibinfo{author}{\bibfnamefont{P.}~\bibnamefont{Hawrylak}},
  \bibinfo{journal}{Phys. Rev. B} \textbf{\bibinfo{volume}{44}},
  \bibinfo{pages}{3821} (\bibinfo{year}{1991}).

\bibitem[{\citenamefont{Pardee and Mahan}(1973)}]{pardee73}
\bibinfo{author}{\bibfnamefont{W.~J.} \bibnamefont{Pardee}} \bibnamefont{and}
  \bibinfo{author}{\bibfnamefont{G.~D.} \bibnamefont{Mahan}},
  \bibinfo{journal}{Phys. Lett. A} \textbf{\bibinfo{volume}{45}},
  \bibinfo{pages}{117} (\bibinfo{year}{1973}).

\bibitem[{\citenamefont{Penn et~al.}(1981)\citenamefont{Penn, Girvin, and
  Mahan}}]{penn81}
\bibinfo{author}{\bibfnamefont{D.~R.} \bibnamefont{Penn}},
  \bibinfo{author}{\bibfnamefont{S.~M.} \bibnamefont{Girvin}},
  \bibnamefont{and} \bibinfo{author}{\bibfnamefont{G.~D.} \bibnamefont{Mahan}},
  \bibinfo{journal}{Phys. Rev. B} \textbf{\bibinfo{volume}{24}},
  \bibinfo{pages}{6971} (\bibinfo{year}{1981}).

\bibitem{anderson67} P.~W. Anderson, Phys. Rev. Lett. {\bf 18},
1049 (1967).

\bibitem[{\citenamefont{Andreani et~al.}(1991)\citenamefont{Andreani, Tassone,
  and Bassani}}]{andreani91}
\bibinfo{author}{\bibfnamefont{L.~C.} \bibnamefont{Andreani}},
  \bibinfo{author}{\bibfnamefont{F.}~\bibnamefont{Tassone}}, \bibnamefont{and}
  \bibinfo{author}{\bibfnamefont{F.}~\bibnamefont{Bassani}},
  \bibinfo{journal}{Solid State Commun.} \textbf{\bibinfo{volume}{77}},
  \bibinfo{pages}{641} (\bibinfo{year}{1991}).

\bibitem[{\citenamefont{Ivchenko}(2005)}]{ivchenko05a}
\bibinfo{author}{\bibfnamefont{E.~L.} \bibnamefont{Ivchenko}},
  \emph{\bibinfo{title}{Optical Spectroscopy of Semiconductor Nanostructures}}
  (\bibinfo{publisher}{Alpha Science}, \bibinfo{year}{2005}).

\bibitem{ivchenkopikus} E.L. Ivchenko, G.E. Pikus, \textit{Superlattices and Other
Heterostructures}, (Springer, 2nd ed., 1997).

\bibitem[{\citenamefont{Panzarini et~al.}(1999)\citenamefont{Panzarini,
  Andreani, Armitage, Baxter, Skolnick, Astratov, Roberts, Kavokin,
  Vladimirova, and M.A.Kaliteevski}}]{panzarini99}
\bibinfo{author}{\bibfnamefont{G.}~\bibnamefont{Panzarini}},
  \bibinfo{author}{\bibfnamefont{L.~C.} \bibnamefont{Andreani}},
  \bibinfo{author}{\bibfnamefont{A.}~\bibnamefont{Armitage}},
  \bibinfo{author}{\bibfnamefont{D.}~\bibnamefont{Baxter}},
  \bibinfo{author}{\bibfnamefont{M.~S.} \bibnamefont{Skolnick}},
  \bibinfo{author}{\bibfnamefont{V.~N.} \bibnamefont{Astratov}},
  \bibinfo{author}{\bibfnamefont{J.~S.} \bibnamefont{Roberts}},
  \bibinfo{author}{\bibfnamefont{A.~V.} \bibnamefont{Kavokin}},
  \bibinfo{author}{\bibfnamefont{M.~R.} \bibnamefont{Vladimirova}},
  \bibnamefont{and} \bibinfo{author}{\bibnamefont{M.A.Kaliteevski}},
  \bibinfo{journal}{Fiz. Tverd. Tela} \textbf{\bibinfo{volume}{41}},
  \bibinfo{pages}{1337} (\bibinfo{year}{1999}).

\bibitem{suris} For review see S. M. Kogan and R. A. Suris,
Zh. Eksp. Teor. Fiz. {\bf 50}, 1279 (1966) [Sov. Phys.—JETP {\bf
23}, 850 (1966)]; I. B. Levinson and E. I. Rashba, Usp. Fiz. Nauk
{\bf 111}, 683 (1973) [Sov. Phys. Usp. {\bf 16}, 892 (1973)].

\bibitem{kyrola} E.~Kyr\"{o}l\"{a}, J. Phys. B: At. Mol. Phys.
{\bf 19}, 1437 (1986).

\end{thebibliography}

\end{document}